# An association between information and communication technology and agriculture knowledge management process in Indian milk co-operatives and non-profit organizations: an empirical analysis

By: Ram Naresh Kumar Vangala, Asim Banerjee, B N Hiremath

DA-IICT, Gandhinagar, India

Abstract

The purpose of this study is to examine the relationship between information and communication technology (ICT) and knowledge management processes (KM process) in Indian milk co-operatives and non-government organizations. Both qualitative and quantitative methods have been adopted in this study. Data were collected using questionnaires from 275 members working in both milk co-operatives and non-profit organizations (NGOs). The analysis and hypotheses testing were implemented using structural equation modeling technique (SEM). The results showed that ICT has significant (at $p = 0.001$) and positive effect on KM processes. The results obtained would help managers to better understand the linkage between ICT and KM processes. They could use the results to improve their ICT (tools and infrastructure) for improving the efficiency of KM process in their organizations. The proposed set of metrics could be used as common tools to measure the performance of ICT in KM process in agriculture organizations and for future research.

**Keywords:** Information and Communication Technology (ICT), Agriculture, Knowledge management (KM), Agriculture Knowledge Management (AKM), India, Structural Equation Modeling (SEM)

## 1. INTRODUCTION

In the modern era of globalization, knowledge has been recognized as a valuable organizational resource from a strategic perspective (James, 2004) and an important factor for competitive advantage, effective organizational performance and success (Rai, 2011). Hence Knowledge Management (KM) has become one of the foremost agendas in many organizations, research institutions and academics (Alavi & Leidner, 2001; Tan & Wong, 2015). It is a dynamic and continuous set of the process which enables the organization enhancement and expands their innovation processes (Karadsheh, 2009). American Productivity & Quality Center (APQC) defines KM as "an emerging set of strategies and approaches to create, safeguard, and use knowledge assets (including people and information), which allows knowledge to flow to the right people at the right time so that they can apply these assets to create more value for the enterprise" (Mahmoudsalehi, Moradkhannejad, & Safari, 2012).

Indian agriculture is a complex enterprise which involves millions of small and marginal farmers. Many of these small and marginal farmers are illiterate and have meager resources to access modern technology in agriculture (Yadav, Rasheed Sulaiman, Yaduraju, Balaji, & Prabhakar, 2015). India has been practicing agriculture since ancient times. Hence India has a vast amount of tacit and explicit knowledge in agriculture domain. Therefore KM in agriculture has an immense scope and challenge for managing agricultural knowledge in public, private and non-government organizations in India (Venkatasubramanian & Mahalakshmi, 2012). Agriculture Knowledge Management (AKM) helps in creating knowledge repositories, improve knowledge access, sharing and transfer and enhancing the knowledge environment in rural communities (V.C. Patil, 2011). There are different state and non-state actors like Government, Co-operative sector, Private entities, Non-Government Organizations (NGOs), etc. operating in Indian agricultural sector with different objectives like productivity enhancement, the well-being of the farming community and agri-business opportunities (Gummagolmath & Sharma, 2011). Table 1 provides a description of major activities of different organizations in Indian agriculture.

**Table 1**
An overview of three different entities working in Indian agriculture

| **Government Sector** | **Private Sector** | **Non-Profit Organizations / Co-operatives** |
|---|---|---|
| *Main Objective and Activities* | | |
| - Increase the productive of agriculture crops<br>- Research and development<br>- Education and economic development<br>- Organizing training programs for farmers<br>- Dissemination of knowledge and new technology to farm communities through krishi vigyan kendra (KVK) and agriculture technology management agency (ATMA) | - Production of seeds, fertilizers, pesticides, food processing etc.,<br>- Export and Import<br>- Research and development<br>- Market interventions<br>- Input supply<br>- Commercial and profit marking<br>- Economic development<br>- Disseminating knowledge to farmers | - Involve in community development.<br>- Focus on Extension, farmer groups, self-employment, self-help group<br>- Empowering farm women<br>- Developing leadership quality in rural communities<br>- Transferring technology to farm communities<br>- Providing training programs<br>- Economic, education, socio & cultural development<br>- Disseminating knowledge and creating/gathering local knowledge |

Management of agricultural knowledge takes place at different levels: individual, within communities, within organizations or institutions and networks of them (Engel, 1990). The knowledge for agriculture development is more often than not created, documented or disseminated by one single source or organization (Rafea, 2009). Moreover, different types of organizations produce a different kind of knowledge and the lack of co-ordination or linkage between public, private, agricultural research and extension institutions (Saravanan, 2012) are often cited as a reason for ineffective knowledge transfer to farmers. Hence there is dire need of KM in such agricultural organizations.

The term, information and communication technology (ICT) has been defined differently by many authors. UNDP[1] defined ICT as "the combination of microelectronics, computer hardware and software, telecommunications, and storage of huge amounts of information, and its rapid dissemination through computer networks". According to Michiels and Van Crowder, ICT defined as "a range of electronic technologies which when converged in new configurations are flexible, adaptable, enabling and capable of transforming organizations' and redefining social relations" (Michiels & Van Crowder, 2001). ICT have a prominent role on KM in the organization. It helps in achieving organization effectiveness and to managing its knowledge assets. Various ICT tools will help in capturing, creating knowledge and make it available to the large community (Chadha & Saini, 2014). Knowledge creation, searching, and diffusion can be improved by using ICT, which increases transmission and response speed within the organization (Sher & Lee, 2004). Along with this ICT facilitates storage and sharing of organizational knowledge (Davenport, David, & Beers, 1998; Demarest, 1997; Nonaka, 1995).

ICT can make Indian AKM more substantive by providing affordable, relevant, searchable and up-to-date agriculture information service to farm communities (V.C. Patil, 2011). It supports farmers' to access timely and relevant information, as well as empower the creation and sharing of knowledge of the farming community itself (Mathur & Goyal, 2014). The use of ICT in AKM include community radio, short message service (SMS) and voice-based cellular telephony, information through telecenters, internets kiosks, village knowledge centers etc. are used to transform/support the traditional agriculture extension system (Mittal, 2012; Sulaiman, 2012). Table 2 summarizes categorization of ICT initiatives in Indian agriculture.

**Table 2**
Categorization of ICT initiatives in Indian agriculture

| Name of the Project | Ownership | Contribute by |
|---|---|---|
| *Web-based Technology* | | |
| Agropedia, Rice Knowledge Management Portal (RKMP), AgriTech, KISSAN Kerala, AGRISNET, AGMARKNET, eKirshi | Public | KVKs, State Agricultural Universities, Research Institutes |
| iKisan | Private | Subject Experts |
| aAQUA | Public, Consortium | Subject Experts |
| Electronic solution against agriculture pest (e-SAP) | Public and Private | Subject Experts |
| *Intermediator between user and service provider* | | |
| e-Sagu, Arik | Public | Subject Experts |
| e-Choupal, Tata Kisan Sansar | Private | Subject Experts |
| Digital Green | NGO | Farmers, subject experts |
| MSSRFVKC | NGO | Subject Experts |

---
[1] http://hdr.undp.org/en/content/human-development-report-2001 (21st July 2016)

| *Mobile Technology/Telecommunication* | | |
| --- | --- | --- |
| Kissan Call Center | Public | Subject Experts |
| IFFCO-IKSL | Public and Private | Subject Experts |
| RML, mKrishi, Nokia Life Tool | Private | Subject Experts |
| Spoken Web | Private and NGO | Subject Experts |
| Fisher Friend Project, Lifelines | NGO | Subject Experts |

Various ICT tools have been deployed for agriculture knowledge management which includes organizational web portals created for specific commodities, sectors, and enterprise and for e-commerce activities (Sulaiman, 2012). A careful analysis of these websites and portals indicates that these are mostly used for disseminating generic information and poor in quality (Balaji, 2009; Yadav, et al., 2015). An electronic database like audio and video recordings and multimedia presentation are widely used for disseminating knowledge. E-mails and discussions forms are commonly used to share knowledge among subject experts, research group and professionals in organizations (Sulaiman, 2012). Portals like Agropedia, RKMP, and Digital Green are developing an agricultural knowledge repository in the form of audio and video visual encyclopedia. Analysis of these ICT projects (Table 2) in Indian AKM revealed that they primarily focus on the transfer of knowledge to farm communities, following a one-way flow of knowledge i.e. from experts to farmers without many opportunities for interaction. Many ICT projects are pushing external content towards local people based on what experts think the community needs (Glendenning & Ficarelli, 2012). Researchers and subject experts still following the pattern of transfer-of-technology, based on assumption that knowledge is created by scientists, subject experts to be packaged and spread by extension officers and to be adopted by farm communities (Assefa, Waters-Bayer, Fincham, & Mudahara, 2009; Waters-Bayer & Van Veldhuizen, 2004). ICT in AKM has been mostly used to support traditional extension system (Sulaiman, 2012). Hence there is a need to focus on how ICT affect knowledge management process (acquiring, creating, storing, organizing, and sharing or disseminating) at the organization level for effective AKM.

Researchers have mostly focused on the influence or impact of ICT in Indian agriculture. There are very limited studies on knowledge management process at the organizational level and still very few on the relationship between ICT and knowledge management process at agricultural institutional or organizational levels in the Indian context. It is not clear how the relation between ICT competency and knowledge management process works. Empirical work in this area is lacking. In this paper, we are trying to address the relationship between ICT and agriculture knowledge management process in Indian agricultural organizations by using structural equation modeling (SEM).

The research paper is arranged as follows. In section 2 we brief on research framework and hypotheses used in this study. Section 3 is followed by a description of the methodology used for conducting a survey. The next section 4 presents the data analysis and the results of data analysis and hypotheses testing. A discussion of the overall results and implications of the study follows in Section 5. The paper culminates with conclusions, together with limitations of the study and future research directions provided in Section 6.

## 2. RESEARCH FRAMEWORK AND HYPOTHESES

As mentioned earlier, very little studies have analyzed the relationship between knowledge management process and ICT in Indian agricultural organizations. Therefore, this research aims to discover the linkage between these two aspects. The main objective of this study is to understand the relations knowledge enabler like ICT and knowledge management process. In this study, ICT is an independent variable and knowledge management process as the dependent variable. Figure 1 is proposed research framework.

### 2.1 Knowledge Management process (KMP)

Various studies have addressed the knowledge management process; they divide the knowledge management into many processes. KM process includes activities of acquiring, creating, storing, sharing, diffusing, developing and deploying knowledge by individuals and groups (Demarest, 1997; Zheng, Yang, & McLean, 2010). According to Davenport and Prusak, KM had three processes that have received that most consensus: knowledge generation, sharing and utilization (Davenport & Prusak, 1998). Alavi and Leidner proposed four processes such as creation, storage, transfer and application (Alavi & Leidner, 2001). Bhatt considered creation, validation, presentation, distribution and application (Bhatt, 2001). Many frameworks for knowledge management process have been identified. This study examines four processes: acquiring and creating, organizing and storing, sharing or disseminating and applying as proposed by Vanagla *et al*, for the agricultural organization (Vangala, Hiremath, & Banerjee, 2014).

#### 2.1.1 Knowledge acquiring and creating (KAC)

In terms of processes, knowledge acquiring and creating is where members in the organization gain, collect, create and obtain required and useful knowledge to perform their job activities. It is a complex, multidimensional and dynamic process. KAC involves developing new content and updating existing content with the organization's tacit and explicit knowledge (Pentland, 1995). Knowledge creation is recognized as the process where new ideas, best practices are generated (Morey, 2001). It is about obtaining knowledge from external/internal sources or the recovery of the knowledge (explicit or tacit) that resides inside the people working in the organization (Jackson, 2001; Supyuenyong & Islam, 2006). For creating new knowledge, it requires everyone in the organization to work in teams and be involved in a non-stop process of personal and organizational self-renewal (Nonaka, 1991). Creation of knowledge in an organization involves a continuous interplay between tacit and explicit knowledge and it develops into the spiral flow as knowledge moves through individuals and groups at different organizational levels (Alavi & Leidner, 2001). According to Nonaka, knowledge creation takes place in four modes within an organization: socialization (tacit to tacit), externalization (tacit to explicit), combination (explicit to explicit) and internalization (explicit to tacit) (Nonaka, 1994). Training programs, workshops or seminars are another means for employees/members of the organization for acquiring and creating new knowledge (Chen, Duan, Edwards, & Lehaney, 2006; M. R. Lee & Lan, 2011). Members or employees of organization rely on technology like the internet for acquiring work-related knowledge to perform their daily work (M. R. Lee & Lan, 2011).

*2.1.2  Knowledge organizing and storing (KOS)*

This process consists of codifying, storing, refining, indexing, evaluating and updating the knowledge in an organization's repository (Rollett, 2012). Knowledge is validated, codified to put in the useful format before it can be used (Tan & Wong, 2015). Once it is evaluated, it is categorized and represented in a structured manner with indexing/mapping to facilitate efficient storage in the organization's repository and its effective usage at a later point (Nonaka, 1995; Rollett, 2012). By updating the existing stored knowledge reduce redundancy, improve quality and minimizes obsolescence (Davenport & Klahr, 1998). Therefore for efficient storage in the repository, it should be archived periodically to provide backup that can be used in case of failure or crash of the machines/servers (Rowley, 2001). ICT tools such as electronic document management, document information systems, and document imaging systems etc., are used for storing knowledge in the organization (Hendriks, 1999).

*2.1.3  Knowledge sharing/disseminating (KSD)*

It is processed by which sharing of knowledge take place among individuals and/or groups in the organization, thereby promoting learning and creation of new knowledge. Knowledge sharing is where tacit and explicit knowledge is disseminated throughout the whole organization (Tan & Wong, 2015). It is considered as a core process of KM since one of the main goals and objectives of KM is to promote sharing of knowledge among individuals, groups and organizations (Chua, 2004; Karadsheh, 2009). Transfer of knowledge can be in the horizontal and/or vertical directions. Horizontal knowledge transfer takes place between the employees in the organizations and vertical knowledge transfer takes place between organizations. Knowledge sharing process can be driven by a formal, informal and personal approach such as meetings, discussions, social network, collaboration, focus group meetings, face-to-face interaction (Marwick, 2001; Tan & Wong, 2015). Knowledge in the organization transferred through social networks, collaboration, and daily interaction whereby chatting and conversation (Davenport & Prusak, 1998). According to Choo, in organizations, members combine their explicit knowledge by sharing/exchanging reports, memos and a variety form of other documents (Choo, 1996). ICT tools, such as e-mails, groupware, networking tools and others can support and boost for effective knowledge sharing (de Carvalho, 2001; Tan & Wong, 2015).

*2.1.4  Knowledge Applying (KAP)*

Knowledge applying is to make good use of knowledge and members/employees of an organization can apply and adopt the best practices in their daily work (O'Dell & Essaides, 1998). This process also means to put knowledge into practice, where employee should apply lessons learnt from previous experience or mistake (Datta, 2007). According to Davenport and Klahr, the effective application of knowledge can assist the organization to improve efficiency and reduce cost (Davenport & Klahr, 1998). Knowledge application includes the application of decision-making protection, action and problem solving which final lead to knowledge creation (Allameh & Zare, 2011).

**2.2     Knowledge management enabling factors**

Knowledge enables are characterized as influencing factors, facilitate knowledge management activities as codifying and sharing knowledge assets among individual and group in the organization (Chan & Chau, 2005). Enabling factors like organization culture, organization structure, technology have the influence on knowledge management in the organization. These enablers are the tools for an organization to develop its knowledge and motivate to create, share and protect knowledge within the organization (Zheng, et al., 2010). A variety of knowledge management enablers have been addressed in the literature (Akbari, Saeidipour, & Baharestan, 2013; Allameh & Zare, 2011; H. Lee & Choi, 2003; Mahmoudsalehi, et al., 2012; Zheng, et al., 2010). Among them, information and communication technology (ICT) has been incorporated in this study.

*2.2.1  Information and communication technology (ICT)*

ICT plays an important role in facilitating communication that often inhibits the interaction between different parts of the organization (Allameh & Zare, 2011). Many researchers have found that ICT is a crucial element for knowledge management process (Alavi & Leidner, 2001; Davenport & Prusak, 1998; H. Lee & Choi, 2003). It supports communication, collaboration, knowledge seeking and enables collaborative learning (Ngoc, 2005). ICT tools help in capturing knowledge and expertise created by knowledge workers and making it available to the large community (Chadha & Saini, 2014). Information technology is widely used in an organization, and thus qualifies as a natural medium for the flow of knowledge in the organization (Allameh & Zare, 2011). Thus, we hypothesize:

*H1: There is positive relationship between ICT and knowledge management process*

All the measurement items and their constructs are listed in Appendix I.

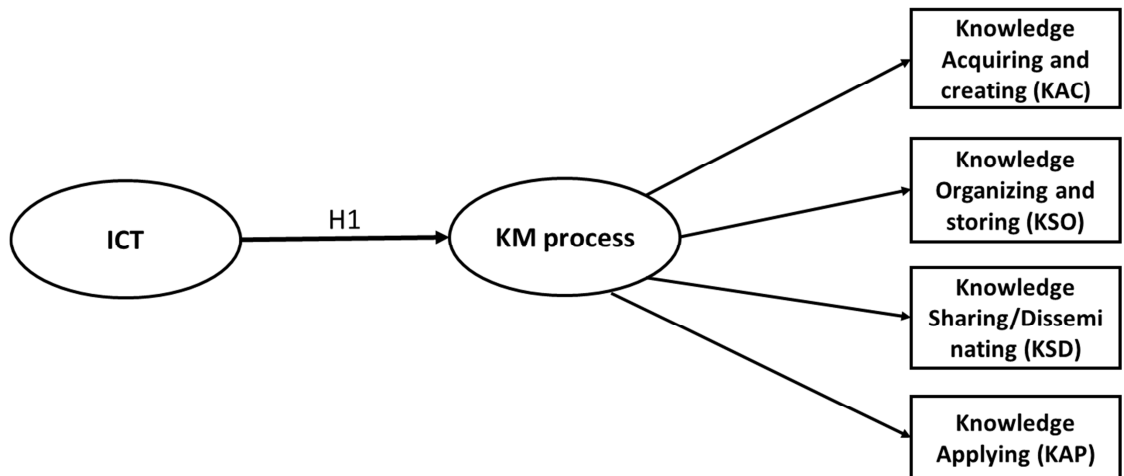

**Figure 1**: Research framework

## 3. RESEARCH METHODOLOGY AND DATA COLLECTION

Both qualitative and quantitative research approach were used to empirically test the research hypothesis. Semi-structured and group discussion were conducted to get a deeper understanding of the flow of knowledge in the organization and the challenges employees faced in using ICT in their knowledge mobilization activities in the organizations. A survey questionnaire was designed to determine and understand the linkage between ICT and KM process. The questionnaire was split into two main sections. The first section includes the general information such as name, gender, position, education and number of years of experience in the organization. The second section investigated the critical metrics for measuring ICT and KM process (acquiring and creating, organizing and storing, sharing/disseminating and applying) that were derived from the literature (Choy, Yew, & Lin, 2006; Daghfous & Kah, 2006; Gold & Arvind Malhotra, 2001; H. Lee & Choi, 2003; Y. C. Lee & Lee, 2007). Respondents were asked to rate the extent to which the metric were practiced in their organization using a five-point scale (from 1 = strongly disagree to 5 = strongly agree).

The authors' selected two milk co-operatives and two non-government organizations that exhibited a strong desire for mobilization and disseminating knowledge to the farm communities using ICT (see Table 3 provide case study details). For each case, an employee from multiple departments participated in filling questionnaires and semi-structure interview or group discussion. Unit of analysis in this study is middle-level managers, veterinary doctors, agriculture extension officers, project coordinators, cluster in-charge or supervisor and field workers/operators. These people are surveyed because they play a key role in managing knowledge. These people are positioned at the intersection of both vertical and horizontal flow of knowledge. Therefore they can synthesize the tacit knowledge of both top (scientist group) and bottom (farmer group) level, convert them explicit knowledge, and incorporate it into the organizational knowledge repository. There is no personal or formal relationship between researchers and interviewees or the organization as a whole. This allowed for triangulation and also helped to validate data interpretation and findings (Venkitachalam & Bosua, 2014).

### 3.1 Data collection

Before running an actual survey, the questionnaire has gone through the pilot test, to ensure the objectives of the questionnaire are clear. The questions are well-structured, understandable and were developed in four languages namely English, Hindi, Gujarati and Telugu because the composition of people working and geographical location of milk co-operatives and NGOs that were the part of the study. A total of 283 respondents were collected from the four organizations. Some of these respondents were also interviewed (by semi-structure and group) to get a deeper understanding of the flow and management of knowledge and challenges faced by the members/employees in their knowledge mobilization activities. Data was collected during their weekly and monthly meetings in the organization. During the meetings, questionnaires were distributed to participants and they were asked to fill the form. Before filling the form, the objectives of the research and questionnaire were explained to them.

**Table 3**
Summary of cases in terms of mission and vision, services and operation

| Organization name | Location | Mission and vision, services and operations |
|---|---|---|
| Mulukanoor Women's Cooperative Dairy | Telangana | - To improve the overall quality of life to dairy producers & consumers by running a sustainable self-sufficient and managed women cooperative |

| (MWCD) | | union, setting an example for collective action and rural women capacity.<br>- To provides services like animal health, feed and fodder, technical inputs, milk marketing, creating value-added milk products and marketing, co-operative development, self-help groups etc., |
|---|---|---|
| Mehsana District Co-operative Milk Producers' Union Ltd (MDCM) | Gujarat | - Enhancing the milk production capacity, providing self-employment and sustainable income generation to the rural farmers, enhancing the per capita availability of milk etc.<br>- To provides services like animal health and nutrition, feeding and health care management, technical inputs, co-operative development and services, milk marketing, creating value-added milk products and marketing etc., |
| Dhruva (DHRU) | Gujarat | - To mobilize, inspire and enable the tribal people through a participatory approach to working towards their own rehabilitation using their own resources which lead to local capacity building and sustainable improvement in their livelihood and quality of life.<br>- To support and provide livelihood generation to tribal people through farming system improvement, watershed development, livestock management, women's development, health, sanitation and nutrition, micro-finance, agro-produce processing and marketing and strengthening of local communities and user groups through the formation of People's Organizations'. |
| Digital Green (DG) | Telangana, Andhra Pradesh | - To integrate innovative technology with global development efforts to improve human well-being<br>- Engage with and empower rural communities to produce participatory localized videos, leveraging pre-existing group structures to disseminate these videos through human mediation. These videos are of the community, by the community, and for the community. |

## 4. DATA ANALYSIS AND RESULTS

The data analysis is the most crucial part of this study as it is used to make inferences about the use of ICTs in KM processes in NGOs and milk co-operatives. The data analysis for collected data was performed by using Statistical Package for the Social Sciences (SPSS version 20.0). Further analysis was conducted by using structural equation modeling (SEM) via the Analysis of Moment Structures (AMOS version 20.0) software. SEM is a multivariate statistical analysis technique that is used to analyze structural relationships. It is the combination of factor analysis and multiple regression analysis. This method is preferred because it estimates the multiple and interrelated dependence in a single analysis[2]. In this study, the analysis was divided into three parts, which were the first-order confirmatory factor analysis (CFA) and second-order CFA for the measurement model and the structural model analysis

## 4.1    Demographic Profile of Respondents

A sample of 283 respondents was collected from the four organizations. Out of 283 responses, 8 responses were invalid as the questionnaire was not completed. There 275 responses were found usable. Table 4 summarizes the profile of the respondents.

**Table 4**
Demographic profile of respondents

| Sample characteristics | Frequency (n=275) | Percent (%) |
|---|---|---|
| *Gender* | | |
| Male | 184 | 67.0 |
| Female | 91 | 33.0 |
| *Education* | | |
| High school | 97 | 35.3 |
| Bachelor Degree | 137 | 49.8 |
| Master Degree | 41 | 14.9 |
| *Working position of respondents* | | |
| Managers | 8 | 3.0 |
| Project in-charge / Program managers | 40 | 14.5 |
| Veterinary doctors / Agricultural Officers | 47 | 17.1 |
| Field in-charge/Supervisor | 180 | 65.4 |
| *Experiences of respondents* | | |
| 0 – 5 years | 101 | 36.7 |
| 6 – 10 years | 96 | 35.0 |
| 11 – 20 years | 55 | 20.0 |
| Above 20 years | 23 | 8.3 |

## 4.2    Assessment of Reliability and Validity Test

Choi has emphasized the importance of both reliability and validity in the data collection and instrument development stages (Choi, 2010). The term reliability refers to the consistency of a research study or the degree to which an assessment tool produces stable and consistent results. For testing of reliability, internal consistency method is used in this study. Internal consistency reliability estimates relate to item homogeneity or the degree to which the items on a test jointly measure the same concept or construct (Henson, 2001) and hence it is connected to the inter-relatedness of the items within the test (Mohsen Tavakol, 2011). Cronbach's alpha, one of the most commonly used coefficient methods to assess the internal consistency of the items (Sekaran, 2006). Alpha is expressed as a number between 0 and 1. It suggests that as a rule of thumb, a Cronbach's alpha value of greater than or equal to 0.7 is required to satisfy the internal consistency reliability (Hair, 2006). Referring to Table 5, this condition has been satisfied for all the constructs.

---
[2] http://www.statisticssolutions.com/structural-equation-modeling/ (accessed on 2nd September 2016)

Validity is defined as the degree to which a measurement assesses what it is supposed to measure. Convergent and discriminant validity were checked for each construct in this study to test validity. Convergent validity refers to the degree to which items that should be related are in reality related, while discriminant validity signifies the degree to which items that should not be related are in fact not related (Tan & Wong, 2015). For convergent validity, the composite reliability (CR) value must be greater than or equal to 0.7 and the average variance extracted (AVE) value must be greater than or equal to 0.5 (Hair, 2006; Segars, 1997). As shown in Table 5, all the constructs have fulfilled these two requirements.

Unidimensionality is achieved when the items have acceptable factor loadings must be greater than or equal to 0.5 (Hair, 2006; H. Lee & Choi, 2003). During the validation process, 2 items (ICT2, ICT4) from independent variable and 5 items (KAC5, KAC6, KOS1, KOS2, KSD9) from dependent variable were dropped due to poor factor loading of less than 0.50. The results of unidimensionality for all the constructs are shown in Table 5. Discriminant validity refers to the degree to which measures of different concepts are different. It is used because each variable was measured by multi-items. Discriminant validity is achieved when the square root AVE for the each construct is higher than the correlation coefficients among the constructs (Hair, 2006). Referring to Table 6, this condition has been satisfied.

**Table 5**
Result of unidimensionality, reliability, convergent validity and discriminant validity.

| Constructs | No. of Items | Indicators | Factor loadings | CR (>=0.7) | AVE (>=0.5) | Cronbach's alpha |
|---|---|---|---|---|---|---|
| Information Communication Technology (ICT) | 5 | ICT6 | 0.849 | 0.836 | 0.561 | 0.791 |
| | | ICT3 | 0.749 | | | |
| | | ICT1 | 0.755 | | | |
| | | ICT5 | 0.691 | | | |
| | | ICT7 | 0.640 | | | |
| Knowledge acquiring and creating (KAC) | 4 | KAC3 | 0.789 | 0.854 | 0.532 | 0.700 |
| | | KAC4 | 0.759 | | | |
| | | KAC2 | 0.723 | | | |
| | | KAC1 | 0.637 | | | |

| Construct | | | | | | |
|---|---|---|---|---|---|---|
| Knowledge organizing and storing (KOS) | 4 | KOS4 | 0.823 | 0.845 | 0.600 | 0.771 |
| | | KOS6 | 0.790 | | | |
| | | KOS5 | 0.772 | | | |
| | | KOS3 | 0.694 | | | |
| Knowledge sharing and disseminating (KSD) | 8 | KSD8 | 0.812 | 0.941 | 0.550 | 0.810 |
| | | KSD7 | 0.779 | | | |
| | | KSD2 | 0.771 | | | |
| | | KSD1 | 0.728 | | | |
| | | KSD4 | 0.721 | | | |
| | | KSD3 | 0.720 | | | |
| | | KSD5 | 0.696 | | | |
| | | KSD6 | 0.694 | | | |
| Knowledge Applying (KAP) | 3 | KAP2 | 0.779 | 0.888 | 0.594 | 0.703 |
| | | KAP3 | 0.777 | | | |
| | | KAP1 | 0.756 | | | |

**Table 6**
Results of discriminant validity analysis

| Construct | ICT | KAC | KOS | KSD | KAP |
|---|---|---|---|---|---|
| ICT | **0.748** | | | | |
| KAC | 0.676 | **0.729** | | | |
| KOS | 0.478 | 0.388 | **0.782** | | |
| KSD | 0.478 | 0.591 | 0.298 | **0.744** | |
| KAP | 0.178 | 0.509 | 0.191 | 0.610 | **0.770** |
| **Note:** The square root of AVE value for each construct is printed along the diagonal, while the correlation coefficient between each pair of construct is presented as the off-diagonal element | | | | | |

Next, the second-order CFA was conducted for the first-order constructs (KAC, KOS, KSD and KAP) of the study. It was used to confirm that underlying measurement constructs loaded into their respective theorized construct (Tan & Wong, 2015). In this respect, the factor loadings between the first-order constructs and second-order constructs must be greater than or equal to 0.5 (Hair, 2006). Referring to Table 7, this condition has been fulfilled and the model of KMP is illustrated in figure 2.

**Table 7**
Second order CFA

| Second order construct | First order constructs | Factor loadings (≥0.5) |
|---|---|---|
| KM process | KAC | 0.798 |
| | KOS | 0.765 |
| | KSD | 0.945 |
| | KAP | 0.807 |

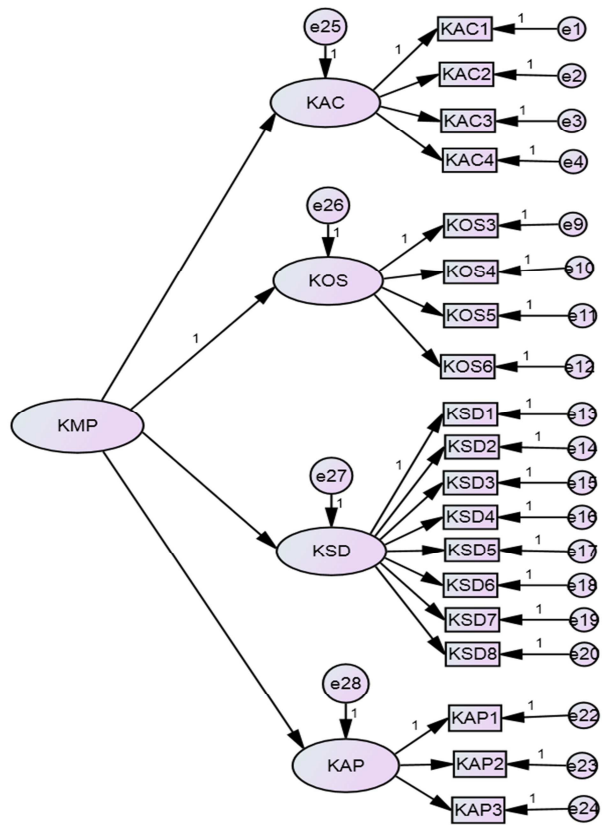

**Figure 2:** KAC, KOS, KSD and KAP constructs in the second-order CFA

### 4.3 Structural Equation Modeling (SEM)

In this study, the structural equation modeling (SEM) was tested using maximum likelihood method. SEM has been widely adopted in social science research using quantitative studies because it permits modification and assessment of theoretical models (Bentler, 1983). It is designed to assess how good a proposed conceptual model can fit the data collected and also to establish the structural relationships between the sets of latent variable (Byrne, 2013). The final model of the study is illustrated in Figure 3. The curved bi-direction arrow (as shown in Figure 3) represents the covariance or correlation between the indicated pair of measurement errors of the respective items due to redundancy. Therefore, the correlated errors were set to be "free parameter estimates" using the double-headed arrow (Byrne, 2013; Tan & Wong, 2015).

To ensure the fitness of the structural model, i.e. how well the data set fits the research model, there are several indicators which are computed by using AMOS. The most fundamental measure of overall fit in a structural equation model is the likelihood-ratio chi-square statistics. As suggested by Bagozzi and Yi, a p-value exceeding 0.05 and a normed chi-square value ($\chi^2$/df) that is below 3, are normally considered as acceptable (Bagozzi & Yi, 1988). Along with this, fitness of the structural model can be studied by using the Comparative Fit Index (CFI) must be greater than or equal to 0.9 (Bentler, 1990), Root Mean Squared Error of Approximation (RMSEA) must be less than or equal to 0.08 (Browne & Cudeck, 1992), Goodness-of-Fit Index (GFI) must be greater than or equal to 0.9 (Hair, 2006) and Adjust Goodness-of-Fit Index (AGFI) must be more than or equal to 0.9 (Hair, 2006). The developed model has been proven to meet all the requirements and the results are shown in Table 8. Hence, the model was utilized to test the hypothesized relationships among the constructs (see Figure 2). Table 9 presents the hypothesis testing result for the causal effect of ICT on KM process. The results revealed that ICT has a significant and direct effect on KM process. Therefore *H1* was supported and accepted.

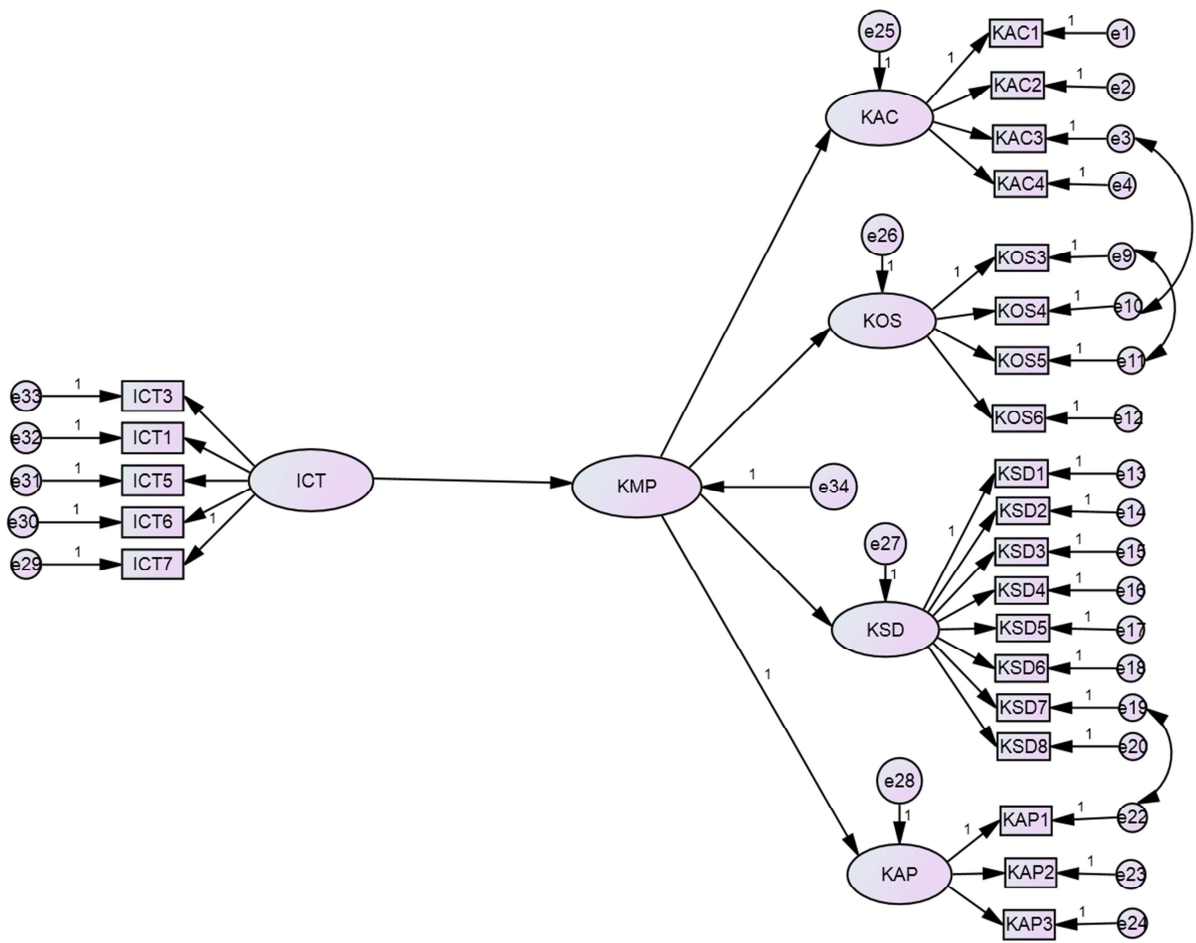

**Figure 3:** Finalized model of the study

**Table 8**
Goodness of fit statistics indicators

| Name of the index | Value obtained | Level of Accepted Fit | Results |
|---|---|---|---|
| chi-square value ($\chi^2$/df) (Chi-square = 662.7 Degrees of freedom = 242) | 2.738 | Below 3 and p=0.001 | Acceptable |
| CFI | 0.931 | >=0.90 | Acceptable |
| RMSEA | 0.081 | =< 0.08 | Acceptable |
| GFI | 0.917 | >=0.90 | Acceptable |
| AGFI | 0.907 | >=0.90 | Acceptable |

**Table 9**
Fitness indexes of the overall model

| Hypothesis | Beta value | p-value | Comment |
|---|---|---|---|
| H1: ICT → KM process | 0.44 | *** | Significant |
| Note: *** significant at 0.001 | | | |

## 5. DISCUSSION AND LIMITATIONS

In this study, the SEM approach was applied to examine the relationship between ICT and KM process in Indian agricultural organizations. As evident from the analysis conducted above, ICT was found to have a significant effect on KM process in the respondent organizations in India. This is in agreement with the proposition of Alavi & Leider that information technology can lead to a greater breadth and depth of knowledge creation, storage, transfer and application in organizations (Alavi & Leidner, 2001). The result is also consistent with the findings from past studies. For instance, Chadha *et al.* found that ICT enhances the visibility of knowledge and facilitate the process of acquiring, creating, storing and disseminating (Chadha & Saini, 2014). Allahawiah *et al.* also verified that there is the positive impact of information technology on knowledge management processes (Allahawiah, Al-Mobaideen, & al Nawaiseh, 2013).

In four of the case organizations, there were clear indications that staffs at various levels and experts have been using Internet, emails for acquiring, storing and sharing knowledge from state and national research institutes. This is substantiated by the statements obtained from various respondents with whom we have interacted during our study. Given below are some excerpts from the interaction we had with them.

Program coordinator [MWCD]: *"Under National Dairy Plan, National Dairy Development Board (NDDB) has provided laptops and the internet connectively for uploading the information of each and every cattle in the village to keep track of ration balance of the cattle."*

Veterinary doctor [MDCM]: *"Under project Ration Balance Program and Productivity Enhance Program, information about all cattle's of the co-operatives were stored in the online database. By assessing this database we can know which village is a shortfall of ration balance, about the Artificial Insemination (AI) requirement and so on. According to that, our doctors prepare their daily route map to visit the villages."*

Program manager [DG]: *"We use the internet to acquire knowledge from experts within and outside the organization"*

Mobile technology is also being widely used for communication and sharing of knowledge with farm communities in all four organizations. Milk co-operatives are using short message services (SMS) for sending alerts on milk procurement, veterinary camps etc. While Digital Green initiated to use interactive voice response (IVR) systems to overcome barriers of literacy. The above is also substantiated by the statements obtained from field supervisors and program coordinators with whom we have interacted during our study. Given below are some excerpts are what they have to say in this regard.

Field supervisor [MDCM]: *"We send SMS to alter to the farmers mobile once milk is procured from them at village collection center. The SMS contain the details of a fat percent, Solid Not Fat (SNF) content and the quantity (liters)".*

Program coordinator [MWCD]: *"Mobile phones have enabled us to quickly contact people in the organization that we think have specific knowledge/information in specified areas to answer specific queries. This, in turn, helps in providing quicker response to farmer query special in the case where I don't have an answer to query".*

Field supervisor [DHRU]: *"Farmers call on my mobile phone to know about pest management for his crop. I use to reply to their quires on the phone itself".*

In Digital Green, digital videos were developed or created on local relevant agriculture and livelihood practices by using ICT tools like video cameras. Then these videos are disseminated by screening for farm communities using battery-operated Pico projectors. All these developed videos are organized and stored in organization repository. These videos can be accessed both offline and online. The above is also substantiated by the statements obtained from field supervisor with whom we have interacted during our study. Given below are some excerpts are what they have to say in this regard.

Field Supervisor [DG]: *"I have been trained by Digital Green in using ICT tools to film/record the best agriculture practices in farm communities. And I disseminate/show this recorded videos to my fellow farmers using Pico projector in village community hall"*

Field Supervisor [DG]: *"We use Pico projects for disseminating agriculture videos to farm communities in offline mode. After screening we collect feedbacks from the farmers, respond to the questions raised by the farmers"*

From the observations and discussions with members/employee, we understand there is a limit of using ICT in organizations. For an instant, we observed that only top and senior management in DHRUVA have access to laptops and internet facilities. The field supervisors use a mobile phone to communicate with peer and farm communities. The above is substantiated by the statement obtained from field supervisor with whom we have interacted during our study. Given below are

Field Supervisor [DHRU]: *"We don't have internet and desktops/laptops with us. Our senior persons have with them. We use mobile phones to disseminate knowledge regarding plant protection, pest management, group meetings, etc., to farm communities. In our daily job we visit farmers' fields personally and interact with them and also attain the calls from them"*

The findings through the analysis of data are consistent with the statement made by various people we interacted during the study and indicate a significant relationship between ICT and KM process.

## 6. CONCLUSION

The easy accesses to ICT and low cost of ICT tools have enhanced development and interest in the field of knowledge management. The availability of ICT has a significant effect on knowledge management process in the case organizations. ICT was found to assist in the process of getting required knowledge and enabling easy communication among the farm communities and organizations. The availability of ICT is seen to enhance dissemination of explicit and tacit knowledge and sharing of best practices effectively among the farm communities and expert groups in the organizations. The rapid developments in the field of ICT for example rapid mobile penetration, availability of the internet, web technologies and mode of communications like emails, video conference etc. helps faster creation, storing, sharing of knowledge within organization. In organizations where face-to-face meetings take very frequently, technology can play a supportive role in recording such meetings for further use.

The results of this study contributed in several ways to the knowledge management theory and practice specific to Indian agriculture. No research of this nature has been conducted in Indian agricultural organizations to assess the relationship between ICT and KM process in agricultural organizations. This study will guide the various levels of managers in selecting of the kinds of tools and technologies to be acquired, with the understanding that lack of support is a major hindrance in the application of technology in KM process in agricultural organizations. The proposed set of metrics could use as common tools to measure the performance of ICT on KM process in agriculture organizations and for future research.

In terms of limitation, the sample used in this study was representing from milk co-operatives and non-profit organizations working in Indian agriculture. Public and private organizations were not covered in this study.

## 7. ACKNOWLEDGEMENT

We would acknowledge all the four organizations Muluknoor Women's Cooperative Dairy, Mehsana District Cooperative Milk Producers' Union Ltd, Digital Green and Dhruva for their cooperation and participation in conducting this study.

**Appendix-I**

| Demographic Profile | |
|---|---|
| **Name of Respondent** | |
| **Name of Organization** | |
| Gender | |
| Male | |
| Female | |
| Education Qualification | |
| High school | |
| Bachelor Degree | |
| Master Degree | |
| Position/Designation in Organization | |
| Managers | |
| Project in-charge / Program managers | |
| Veterinary doctors | |
| Field in-charge/Supervisor | |
| Experience in Organization | |
| 0 – 5 years | |

| 6 – 10 years      |  |
|-------------------|--|
| 11 – 20 years     |  |
| Above 20 years    |  |

| **INFORMATION AND COMMUNICATION TECHNOLOGY (ICT)** | |
|---|---|
| ICT1 | Our organization have ICT infrastructure( like computer, networks) for managing all kind documents on agriculture knowledge |
| ICT2 | ICT infrastructure (like computers, software, networks) are easy to use for uploading, searching and retrieving agriculture knowledge |
| ICT3 | I use ICT tools (like computers, emails, telephones, mobile) to communicate within organization |
| ICT4 | I routinely utilize ICT tools (like computers, emails, telephones, mobile) to access agriculture knowledge from outside organizations |
| ICT5 | We use ICT tools (like computers, emails, telephones, mobile) for sharing agriculture knowledge with farm communities |
| ICT6 | We use computers for storing agriculture knowledge |
| ICT7 | We use internet, intranet to access agriculture knowledge repository |
| **KNOWLEDGE ACQUIRING AND CREATING** | |
| KAC1 | Organization had processes of acquiring agriculture knowledge by collaborating with research institutes, business partners, farm communities |
| KAC2 | Organization give importance's on creating new agriculture knowledge |
| KAC3 | Organization creates manuals and documents on best practices, success stories in agriculture |
| KAC4 | Organization encourages employee, farm communities to exchanges new ideas between individual and group |
| KAC5 | Organization rewards farmers for generating new knowledge in agriculture practices |
| KAC6 | Organization rewards employee for generating new knowledge in agriculture practices |
| **KNOWLEDGE ORGANIZING AND STORING KNOWLEDGE** | |
| KOS1 | Organization utilizes various print material (such as newsletters, handbooks, annual reports, manuals and etc.,) to store agriculture knowledge |
| KOS2 | Organization utilize audios, videos to store agriculture knowledge |
| KOS3 | Database that gathered agriculture knowledge are available in the organization's repository |
| KOS4 | Organization has good IT infrastructure to store the agriculture knowledge |
| KOS5 | Organization use advance IT tools for filtering, listing, indexing the agriculture knowledge to retrieve |
| KOS6 | Knowledge repository (library) are frequently updated |
| **KNOWLEDGE SHARING/DISSEMINATING KNOWLEDGE** | |
| KSD1 | Periodical annual reports/success stories are made to share with all organization members |
| KSD2 | Periodical meetings/workshops/seminars are held to share about best practices, new technology in agriculture |
| KSD3 | Farm communities are willing to share their experience and knowledge with each other |
| KSD4 | Farm communities are willing to share their experience and knowledge with experts group |
| KSD5 | We share our field experience with peer group in the organization |
| KSD6 | We use ICT tools like mobile, audio and video conference, internet for sharing agriculture knowledge |
| KSD7 | Organization encourages employee to share their knowledge with peer groups and others |
| KSD8 | Organization has resources centers, community hall and forums for sharing agriculture knowledge |
| KSD9 | I believe that sharing agriculture knowledge across groups will yield high benefit |
| **KNOWLEDGE APPLYING KNOWLEDGE** | |
| KAP1 | Farmers apply agriculture knowledge to improve their productivity |
| KAP2 | Farmers take the advantage of new technology to improve their work efficiency |
| KAP3 | Farmers use the knowledge to solve the problems in agriculture |